\documentclass[iop,apj,tighten]{emulateapj}
\pdfoutput=1 
\usepackage{apjfonts} 
\usepackage{amsmath,amstext}
\usepackage{gensymb}
\usepackage{color}
\usepackage[breaklinks,colorlinks,citecolor=blue,linkcolor=magenta]{hyperref} 
\usepackage[all]{hypcap} 
\usepackage{bm}
\newcommand{\msun}{\ensuremath {\rm M}_{\odot}}
\newcommand{\yr}{\ensuremath{\rm yr}}

\newcommand{\Gpc}{\ensuremath{\,\rm Gpc}}
\newcommand{\Mpc}{\ensuremath{\,\rm Mpc}}

\def\mnras{MNRAS}  
\def\na{NewA}  

\shorttitle{Dynamics of BH Binaries}
\shortauthors{Hoang et. al.}

\begin{document}

\title{Black hole mergers in galactic nuclei induced by the Eccentric Kozai-Lidov effect}

\author{Bao-Minh Hoang\altaffilmark{1}, Smadar Naoz\altaffilmark{1,2} \\  Bence Kocsis\altaffilmark{3}, Frederic A.~ Rasio\altaffilmark{4 } and Fani Dosopoulou\altaffilmark{4 } }
\affil{$^1$Department of Physics and Astronomy, University of California, Los Angeles, CA 90095, USA}
\affil{$^2$Mani L. Bhaumik Institute for Theoretical Physics, Department of Physics and Astronomy, UCLA, Los Angeles,
CA 90095}
\affil{$^3$Institute of Physics, E\"otv\"os University, P\'azm\'any P. s. 1/A, Budapest, 1117, Hungary}
\affil{$^4$Center for Interdisciplinary Exploration and Research in Astrophysics (CIERA), Northwestern University, Evanston, IL 60201, USA}

\begin{abstract}
Nuclear star clusters around a central massive black hole (MBH) are expected to be abundant in stellar black hole (BH) remnants and BH-BH binaries. These binaries form a hierarchical triple system with the central MBH and gravitational perturbations from the MBH can cause high-eccentricity excitation in the BH-BH binary orbit. During this process, the eccentricity may approach unity, and the pericenter distance may become sufficiently small that gravitational-wave emission drives the BH-BH binary to merge. In this paper, we construct a simple proof-of-concept model for this process and, specifically, we study the eccentric Kozai-Lidov mechanism in unequal-mass, soft BH-BH binaries. Our model is based on a set of Monte Carlo simulations for BH-BH binaries in galactic nuclei, taking into account quadrupole- and octupole-level secular perturbations, general relativistic precession, and gravitational-wave emission. For a typical steady-state number of BH-BH binaries, our model predicts a total merger rate $\sim 1 - 3$$\Gpc^{-3} \yr^{-1}$, depending on the assumed density profile in the nucleus. Thus, our mechanism could potentially compete with other dynamical formation processes for merging BH--BH binaries, such as interactions of stellar BHs in globular clusters, or in nuclear star clusters without a MBH.
\end{abstract}

\maketitle

\section{Introduction}

Recently, the  Advanced Laser Interferometer Gravitational-Wave Observatory\footnote{\url{http://www.ligo.org/}} (LIGO) has directly detected gravitational waves (GWs) from at least five inspiraling black hole-black hole (BH-BH) binaries in the Universe  \citep{LIGO2,LIGO1,LIGO3,GW170608,2017PhRvL.119n1101A}. With ongoing further improvements to LIGO and the commissioning of additional instruments VIRGO\footnote{\url{http://www.virgo-gw.eu/}} and KAGRA\footnote{\url{http://gwcenter.icrr.u-tokyo.ac.jp/en/}}, hundreds of BH-BH binary sources may be detected within the decade, opening the era of gravitational wave astronomy \citep{LIGOrate2016}. 

The primary astrophysical origin of BH-BH mergers is still under debate.
Merging BH binaries may be produced dynamically in dense star clusters such as globular clusters or nuclear star clusters at the center of galaxies \citep{2000ApJ...528L..17P,2003ApJ...598..419W,OLeary06,Antonini2014ApJ...781...45A,Rodriguez16,2016arXiv160202809O,OKL09,Kocsis+Levin,2016arXiv160203831B,2016arXiv160204226S}. These merging BH binaries may also be a result of isolated binary evolution in the galactic field due to special modes of stellar evolution and evolution in active galactic nuclei \citep{Mandel16,Mandel16b,2016arXiv160204531B,Marchant16}. Some studies also suggested that these LIGO detections are sourced in the first stars \citep{2014MNRAS.442.2963K,2016MNRAS.456.1093K,2016arXiv160305655H,2016arXiv160306921I,2016arXiv160404288D}, cores of massive stars \citep{2013PhRvL.111o1101R,2016ApJ...819L..21L,2016arXiv160300511W}, or dark matter halos comprised of primordial black holes \citep{2016arXiv160300464B,2016arXiv160305234C,2016arXiv160308338S}. 

Here we focus on sources produced in galactic nuclei around massive black holes (MBHs). The large escape speed in nuclear star clusters (NSCs) creates an ideal environment to accumulate a population of stellar-mass BHs. Thus, even stellar-mass BH binaries can survive in this configuration despite of supernova kicks (Lu \& Naoz, in prep.). \citet{Antonini+16} investigated the evolution of binaries in NSCs {\it without MBHs in their centers}, and found a rate of $1.5 \Gpc^{-3}\yr^{-1}$. Furthermore, \citet{OKL09} considered binaries that form due to gravitational wave emission during close encounters between single BHs \citep[see also][]{Kocsis+Levin,Gondan+17}. In this channel, the merger rate is dominated by massive BHs over $25 \msun$, which are delivered to the center by dynamical friction and relax to form steep density cusps near the MBH.\footnote{Mergers of low mass BHs, and neutron stars may be rare in this channel \citep{2013ApJ...777..103T}.}

 \begin{figure*}\label{fig:ICs}
\includegraphics[width=\linewidth]{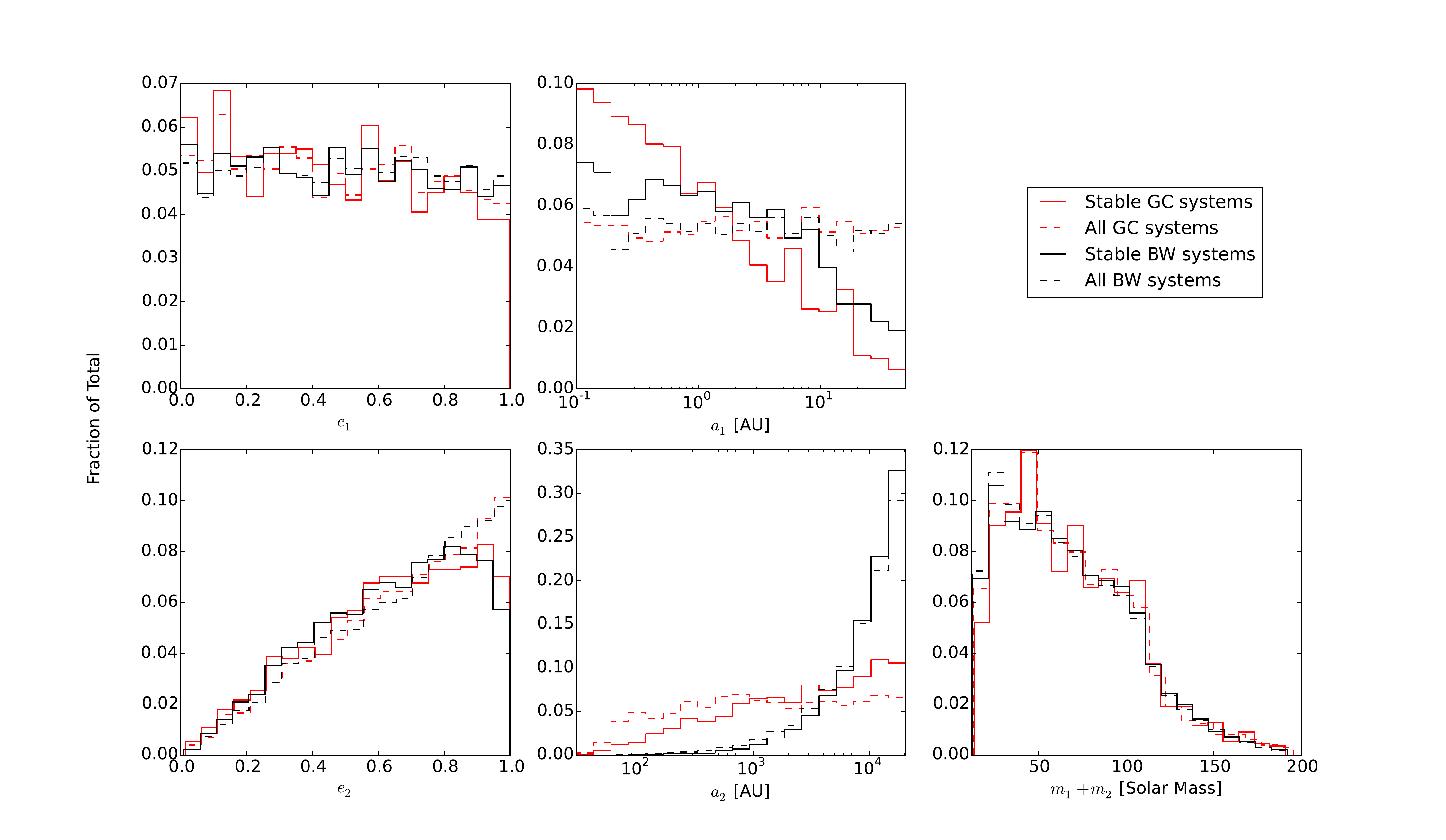}
\caption{The initial conditions distributions before (dashed lines) and after (solid lines) accounting for stability for our nominal runs. We consider the two density profiles we adopt, GC (red lines) and BW (black lines). The top and bottom rows depict the distribution of the inner and outer binaries, respectively, showing $e, a$ from left to right, the bottom right panel shows the initial distribution of the total inner binary mass, $m_1+m_2$.  The mutual inclination is isotropic (uniform in $\cos i$), and the arguments of periapsis $\omega_1,\omega_2$ are uniform from $0$ to $2 \pi$. These distributions did not change after applying the stability criterion and thus are not depicted here to avoid clutter.}
\end{figure*}

We investigate the secular evolution of stellar-mass BH binaries in galactic nuclei {\it which include an MBH in their centers}. These BH-BH binaries undergo large amplitude eccentricity oscillations due to the Eccentric Kozai-Lidov \citep[EKL, e.g.,][]{Naoz16} mechanism in the presence of the MBH \citep[e.g.,][]{Antonini+10,Antonini+12}. If the eccentricity reaches a sufficiently high value, GW emission drives the binary to merge.\citet{Antonini+12} studied the merger rate of BH-BH binaries in the presence of an MBH in galactic nuclei and estimated the merger rates to be $1.7-4.8\times 10^{-4}$~Myr$^{-1}$. \citet{VanLandingham+16} studied the merger rate of BH-BH binaries in the presence of intermediate mass black holes $\sim 10^{3-4}\msun$ and found it to be very high. Here we also focus on BH-BH mergers in the presence of an MBH in galactic nuclei. However, unlike \citet{Antonini+12}, which assumes that the heaviest BH mass in the cluster is $10\,\msun$, we explore a large range of BH masses because recent LIGO observations have shown that there is a much greater range of BH masses than previously thought. EKL effects are stronger when there is a greater mass difference between the two BHs in the binary. Thus, \citet{Antonini+12} concluded that EKL was not important and based their rate estimate on the semi-analytical timescale given in \citet{Tho10}, using only the quadrupole order in a multipole expansion.\footnote{The range in the rate estimation of \citet{Antonini+12} represents core to mass-segregated distributions of the binaries, and corresponds to $\sim 0.002-0.48$~Gpc$^{-3}$~yr$^{-1}$.  }  Here we use the EKL mechanism which represents the secular approximation up to the octupole level of approximation. Furthermore, we adopt a BH population  consistent with a BH mass function that extends to higher masses \citep{OKL09,Kocsis+Levin}.

The octupole correction drives chaotic variations in the eccentricity, increases the probability of close encounters, and thus enhances the efficiency of BH-BH mergers. We show that this mechanism results in a merger rate which is higher than the rates found in the case without MBH, and it is coincidentally comparable to estimated globular cluster rates.

We describe our simulations in Section \ref{sec:ICs}, and present our results, predictions  and merger rate in Section \ref{sec:results}. Finally, we offer our discussions in Section \ref{sec:dis}.

\section{Numerical setup}\label{sec:ICs}

\begin{table}\label{table:simulations}
\begin{center}
\begin{tabular}{| c | c | c | c | c | c |}
\hline
 & {\bf Profile} & {\bf Number of} & {\bf BH mass ($\msun$)} & {$\bf t_{\rm \bf ev}$} & {\bf NT} \\
 & & {\bf Simulations} & & & \\ [0.5ex]
\hline \hline
{\bf Nominal} & GC & 1000 &6-100  & stars & 0 \\
{\bf Runs} & BW & 1500 & 6-100  & stars & 0 \\
\hline
 & BW & 500 & 5-15  & stars & 0 \\
{\bf Additional}  & BW & 1500 & 6-100 & 10 $\msun$ BHs & 0 \\
{\bf Runs} & BW & 1500 &6-100 & 30 $\msun$ BHs & 0 \\
 & GC & 63 & 6-100 & stars & 1 \\
\hline
\end{tabular}
\caption{ Information about the density profile, number of Monte-Carlo simulations performed, BH mass distribution (distribution is for each component of the binary BH), evaporation time ($t_{\rm ev}$, see Equation \ref{eq:evap}), and Newtonian precession (NT), for different simulation runs performed. The first two runs listed are our nominal GC and BW runs. The last four runs listed contain various changes and additions to the two nominal runs. The $t_{\rm ev}$ column denotes whether the evaporation timescale is dominated by the background stars, or background black holes. For the NT column, 1 means that Newtonian precession was included, and 0 means that Newtonian precession was not included.}
\end{center}
\end{table}

We run several large sets of Monte-Carlo simulations to investigate the effects of the SMBH's gravitational perturbations on binary BHs, with variations between sets of simulations to account for different effects (see Table \ref{table:simulations} for details of all simulations).  We study the secular dynamical evolution of binary BHs around the MBH in galactic nuclei, starting from the BH binary phase. We include the secular equations up to the octupole-level of approximation \citep[e.g.,][]{Naoz16}, general relativity precession of the inner and outer orbits \citep[e.g.,][]{Naoz+13GR}, and gravitational wave emission \citep{Peters64}. We consider this as a simple proof of concept to investigate the effects of the EKL mechanism\footnote{Note that we do not assume any specific mechanism for the formation or delivery of those binaries. We later suggest that these binaries might be the result of a continuous stellar formation in the center but we do not exclude other delivery scenarios. If the former is in effect, then in some cases some binaries might merge during the main sequence evolution \citep{Prodan+2015,Stephan+16}}. In general, we neglect the effects of Newtonian precession, which causes the precession of the outer orbit, and thus does not yield a suppression of the EKL mechanism \citep{Li+15}. We have rerun the merged systems via EKL processes for the GC case (see Table \ref{table:simulations}) with Newtonian precession, and demonstrated that it indeed has a negligible effect on the EKL mechanism for the physical picture we considered (see the discussion in Section \ref{sec:additionaltests}).

The number of BHs, their mass distribution, and number density are poorly known in NSCs. Theoretically, a single-mass distribution of objects forms a power law density cusp around a massive object with $n(r)\propto r^{-1.75}$ \citep{Bahcall+Wolf}, where $n(r)$ is the number density and $r$ is the distance from the MBH. For multi-mass distributions, lighter and heavier objects develop shallower ($\propto r^{-1.5}$) and steeper cusps (typically $\propto r^{-2}$ to $r^{-2.2}$, and $r^{-3}$ in extreme cases), respectively \citep{Bahcall+Wolf2,2006ApJ...645L.133H,2006ApJ...649...91F,2009ApJ...698L..64K,Aharon+Perets}. Recent observations of the stellar distribution in the Milky Way NSC identify a cusp with $n(r) \propto r^{-1.25}$ \citep{2017arXiv170103816G,2017arXiv170103817S} consistent with the profile after a Hubble time \citep{2017arXiv170103818B}. As BHs are heavier than typical stars, they are expected to relax into the steeper cusps. The relaxation time of BH populations is much shorter: $0.1$--$1\,$Gyr \citep{OKL09}, although it can become much longer than that in the case of a shallow stellar
density profile \citep{Dosopoulou+Antonini}. In this paper, we assume that the BH number density follows a cusp with either $ n(r) \propto r^{-2}$ or $r^{-3}$ in our two sets of calculations.The BH mass in the two cases is set arbitrarily to $10^7$ and $4\times 10^6\,\msun$, respectively, and we refer to the two models as ``Bahcall-Wolf-like'' (BW) and ``Galactic Center'' (GC) examples.  Note that here ``Galactic Center'' refers to the assumed MBH mass \citep{Ghez+08} and the observed stellar distribution (see below). In reality, the cusp distribution varies with BH mass. Thus, we have also generated initial conditions for the BW case so that $\beta$ in the number density distribution , $n(r) \propto r^{-(3/2 + \beta)}$, is calculated by $\beta(m) = m/4M_0$, where $m$ is the binary mass and $M_0$ is the weight average mass \citep[e.g.][]{2009ApJ...698L..64K, Alexander+Hopman2009,Aharon+Perets}. We found that, due to the stability conditions, the initial condition distribution does not change significantly. For the GC case, the initial conditions distribution will change more significantly if we allow $\beta$ to vary with mass. However, we keep $\beta~=~3$ to investigate the effects of a steep number density distribution on the rates. As we will show later in the paper, the choice of number density distribution will have a very limited effect on the merger rate.

Note that   $dN = 4\pi r^2 n(r) dr = 4 \pi r^3 n(r) d(\ln r)$, where $N$ is the number of objects, and thus we choose to have the initial outer binary semi-major axis follow a uniform distribution in $a_2$ and $\ln a_2$ in the BW and GC model, respectively.  We set the minimum $a_2$ to be one at which the relaxation timescale equals the outer binary gravitational wave merger timescale. {The maximum $a_2$ is chosen to be $0.1$~pc, which corresponds to the value at which the eccentric Kozai-Lidov timescale is equal to the timescale on which accumulated fly-bys from single stars tend to unbind the binary (see Eq.~\ref{eq:evap} and \ref{eq:tquad})}.
However, the BH-binary semi-major axis distribution changes after applying the stability criteria (see below).

\begin{figure*}
\hspace{-3cm}
\centering
\begin{minipage}[b]{.4\textwidth}
\centering
\includegraphics[scale=.45]{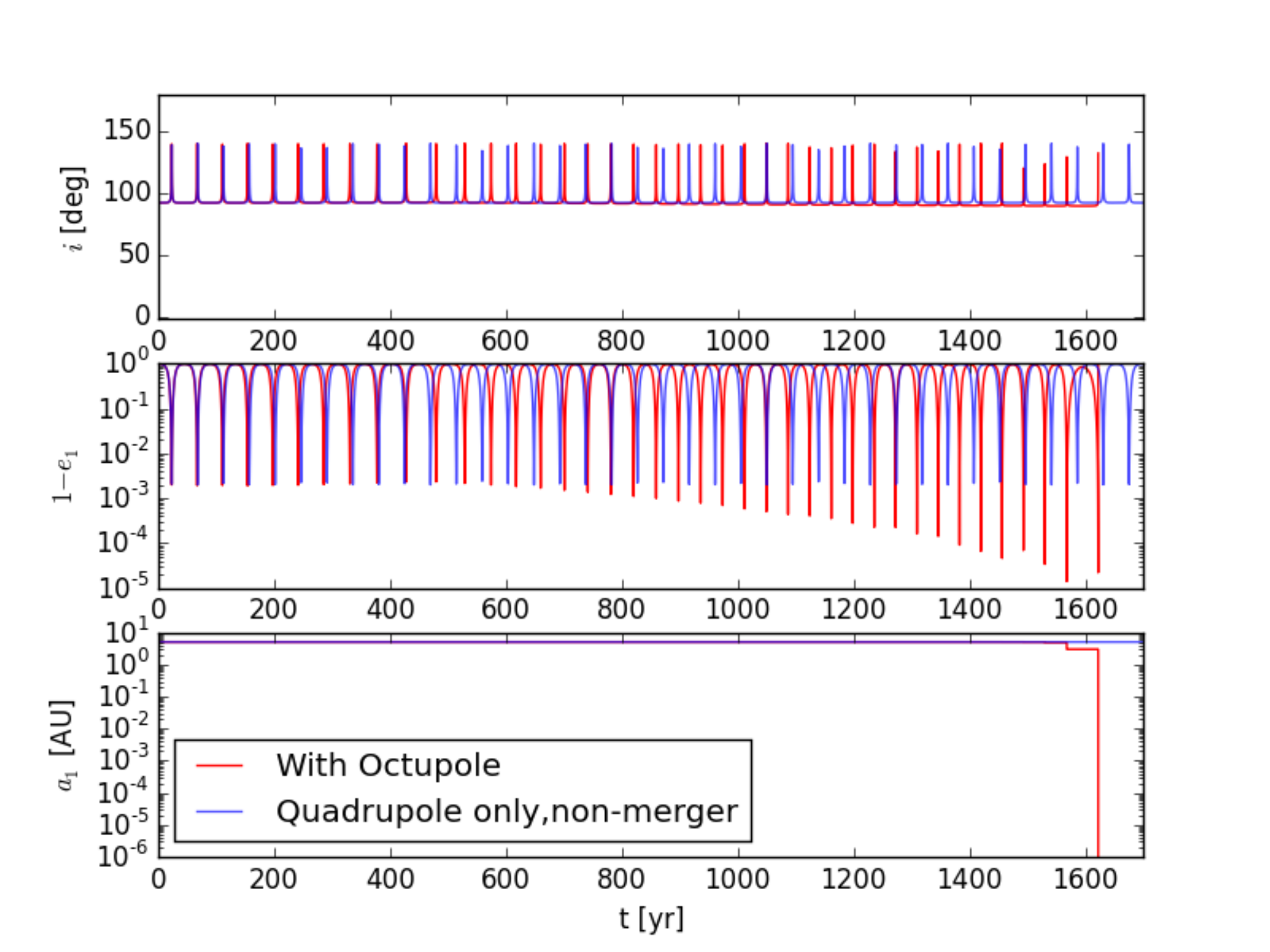}
\end{minipage}\hspace{1cm}
\begin{minipage}[b]{.4\textwidth}
\centering
\includegraphics[scale=.45]{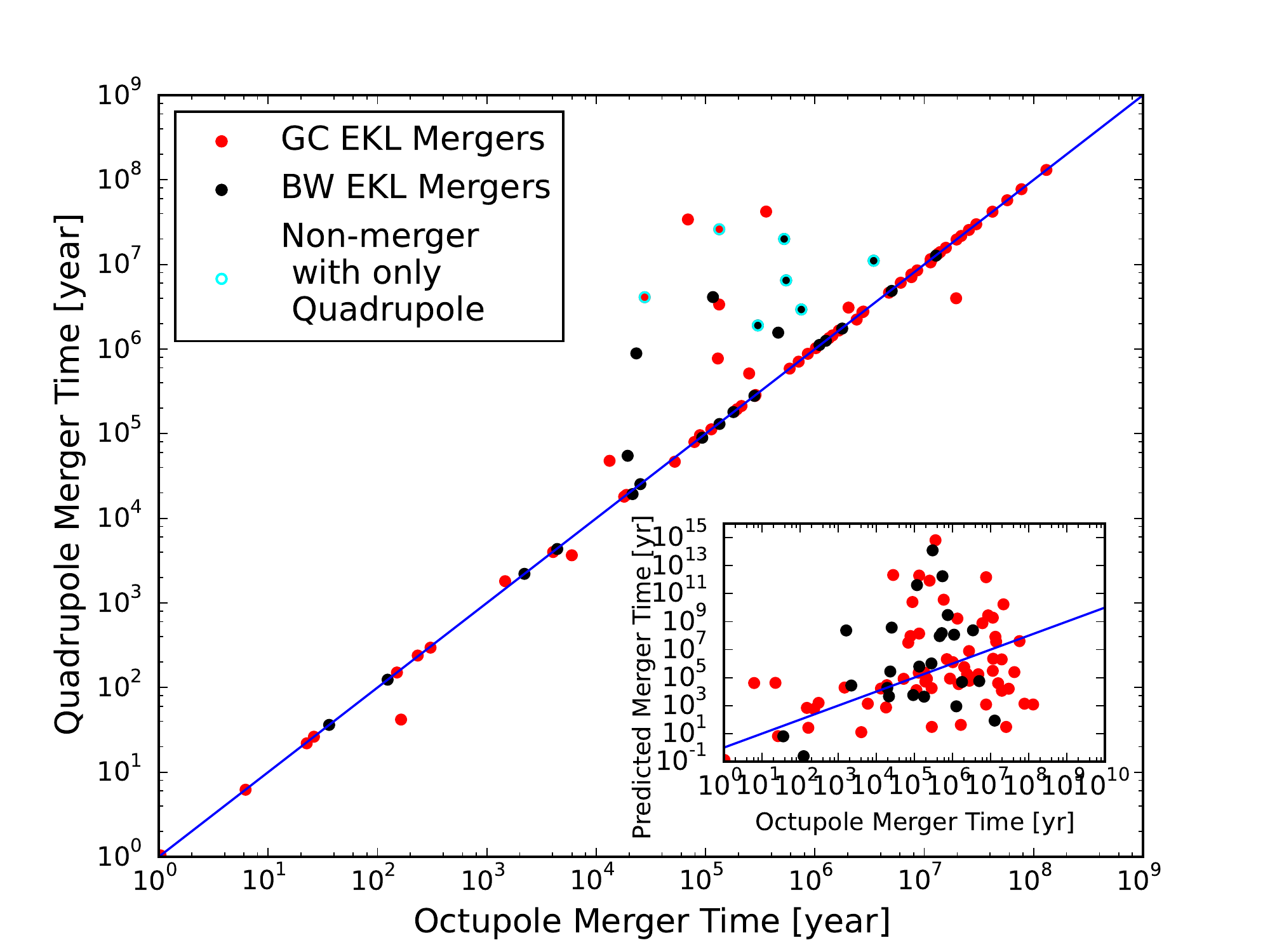}
\end{minipage}
\qquad\caption{{\it Left panel:} Time evolution of the inner binary using up to the octupole level of approximation (red) and only up to the quadrupole level of approximation (blue). The initial conditions are $m_1$ = 12.8 $\msun$, $m_2$ = 63.3 $\msun$, $m_{MBH}$ = $1 \times 10^7$ $\msun$, $a_1$ = 5.1 AU, $a_2$ = 936 AU, $e_1$ = 0.014, $e_2$ = 0.4, $i$ = 92.8 $^{\circ}$. The octupole-level run results in a  merger at 1623 years whereas the quadrupole-level run results in a non-merger. {\it Right panel:}
The merger time at the quadrupole level versus the merger time at the octupole level for the EKL-induced mergers. 
For $16\%$ of the GC systems and $40\%$ of the BW systems the octupole level of approximation is important in predicting the correct merger time. For 2 of the GC systems and 6 of the BW systems, using only the quadrupole level of approximation results in a non-merger.
 These systems are shown with a cyan outline and the timescale shown on the y-axis is $t_{ev}$, (see Eq.~\ref{eq:evap}).  
The inset shows the semi-analytical merger timescale from \citet{Tho10} plotted against the merger time in the Monte-Carlo simulations. 
}
\label{fig:OctvsQuad}
\hspace{1cm}
\end{figure*}

Motivated by the recent LIGO detections, in both examples the mass of each of the BHs is chosen from a distribution uniform in logspace between $6-100$~M$_\odot$ {(i.e. $dN/dm \propto m^{-1}$, \citealt{2002ApJ...581..438M,2004ApJ...611.1080W})}. 
For comparison, we have also run additional Monte-Carlo simulations in the case of the BW distribution, keeping all other parameters the same but using a mass distribution uniform in logspace between $5-15$~M$_\odot$ instead of $6-100$~$_\odot$  \citep{Belczynski+2004}. This had a negligible effect on the EKL induced mergers (see table 1).
The binary separation $a_1$ is drawn from a uniform in log distribution between $0.1-50$~AU.  This is consistent with  \citet{Sana+12} distribution which favors short period binaries, although those BHs have already gone through stellar evolution and thus their exact distribution is unknown. We note that our stability criteria largely modifies the chosen distribution and yields a steeper distribution with a cut-off for systems beyond $50$~AU, as shown in Figure  \ref{fig:ICs} (see below). The lower limit is such as to avoid mass-transfer before the supernova took place. {We note that the vast majority of our binaries are soft, with only $2\%$ ($5\%$) of the GC (BW) binaries being hard binaries \citep[e.g.,][]{Quinlan96}.}

The eccentricity of the BH binaries is chosen from a uniform distribution \citep{Raghavan+10} and taking the outer eccentricity distribution to be thermal \citep{Jeans+1919}.   Furthermore, the inner and outer argument of periapsis $\omega_1$ and $\omega_2$ are chosen from a uniform distribution  between $0$ to $2\pi$. In addition, the mutual inclination $i$ is drawn from an isotropic distribution (uniform in $\cos i$).

After drawing these initial conditions we require that the systems satisfy dynamical stability, such that the hierarchical secular treatment is justified. We use two stability criteria. First we set 
\begin{equation}\label{eq:epsilon}
\epsilon = \frac{a_1}{a_2}\frac{e_2}{1-e_2^2} < 0.1 \ ,
\end{equation}
which  is a measure of the relative strengths of the octupole and quadrupole level of approximations \citep{Naoz16}. Secondly, we require that the inner binary does not cross the Roche limit of the central MBH:
\begin{equation}
\frac{a_2}{a_1} > \Big(\frac{3 m_{\rm MBH}}{m_{\rm binary}}\Big)^{1/3} \  \frac{1+e_1}{1-e_2} \ ,
\end{equation}
\citep[e.g.,][]{Naoz+14DM}\footnote{See \citet{Antonini+12} for quasi-secular evolution.}.
These stability criteria may significantly alter the distribution of BH binary systems 
that can survive to long timescales around the MBH \citep[similar to ][]{Stephan+16}. We show the before and after-stability distributions in Figure \ref{fig:ICs} for both examples. 
We use these distributions to initialize our runs. We note that the distribution of the angles ($\omega_1,\Omega_2$ and $i$) remained the same after applying the stability criteria.

The main difference between the BW and GC models is the stable outer binary semi-major axis ($a_2$) distribution. As depicted in Figure \ref{fig:ICs}, after the stability criteria $\sim 80\%$  of the BW systems remained stable, and their orbital parameter distribution did not significantly change. On the other hand, 
the choice of uniform in $\rm ln(a_2)$ initial distribution for the GC case results in a steeper distribution than the nominal \citet{Bahcall+Wolf} density profile, with $\sim 57\%$ of the systems remaining stable.  Coincidentally, this type of steeper distribution is consistent with the de-projected density profile of the disk of massive stars \citep{Bartko09,Lu09}. Moreover, this distribution represents a strongly mass segregated cluster \citep{2009ApJ...698L..64K}.

\begin{figure}\label{fig:epsi}\hspace{-1cm}
\includegraphics[width=1.2\linewidth]{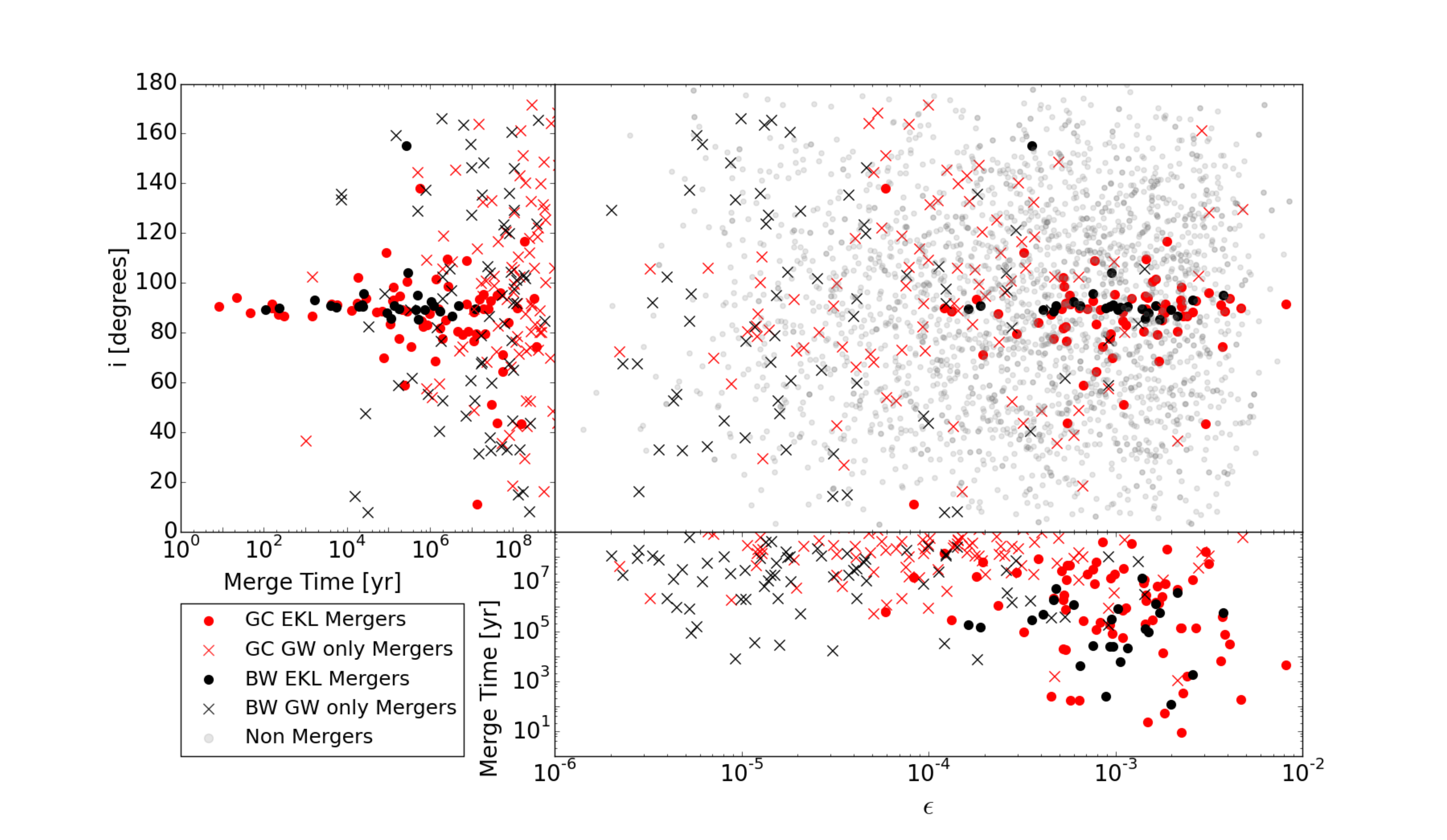}
\caption{{\it Main panel:} The mutual inclination, $i$-$\epsilon$ plane, where $\epsilon$ is the hierarchy parameter (Eq.~\ref{eq:epsilon}). We consider  EKL mergers (red points), GW-only mergers (blue points), and non-mergers (gray points). {\it Left panel:} Distribution of EKL-induced and GW-only mergers for mutual inclination versus merger time. {\it Bottom panel}: Distribution of EKL-induced mergers and GW-only mergers for merger time versus $\epsilon$. 
}
\end{figure}

\begin{figure*}
\hspace{-3cm}
\centering
\begin{minipage}[b]{.4\textwidth}
\centering
\includegraphics[scale=.45]{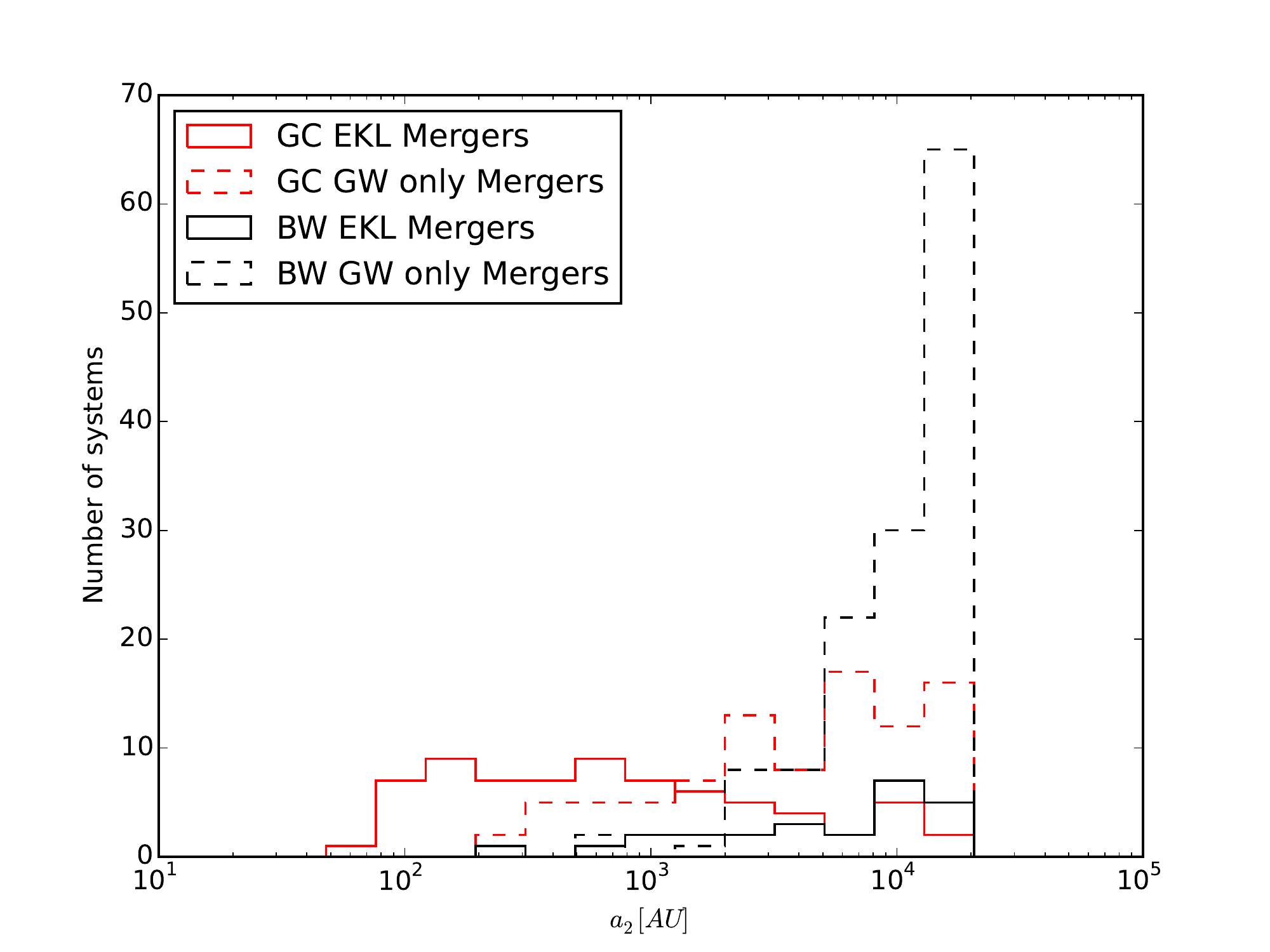}
\end{minipage}\hspace{1cm}
\begin{minipage}[b]{.4\textwidth}
\centering
\includegraphics[scale=.45]{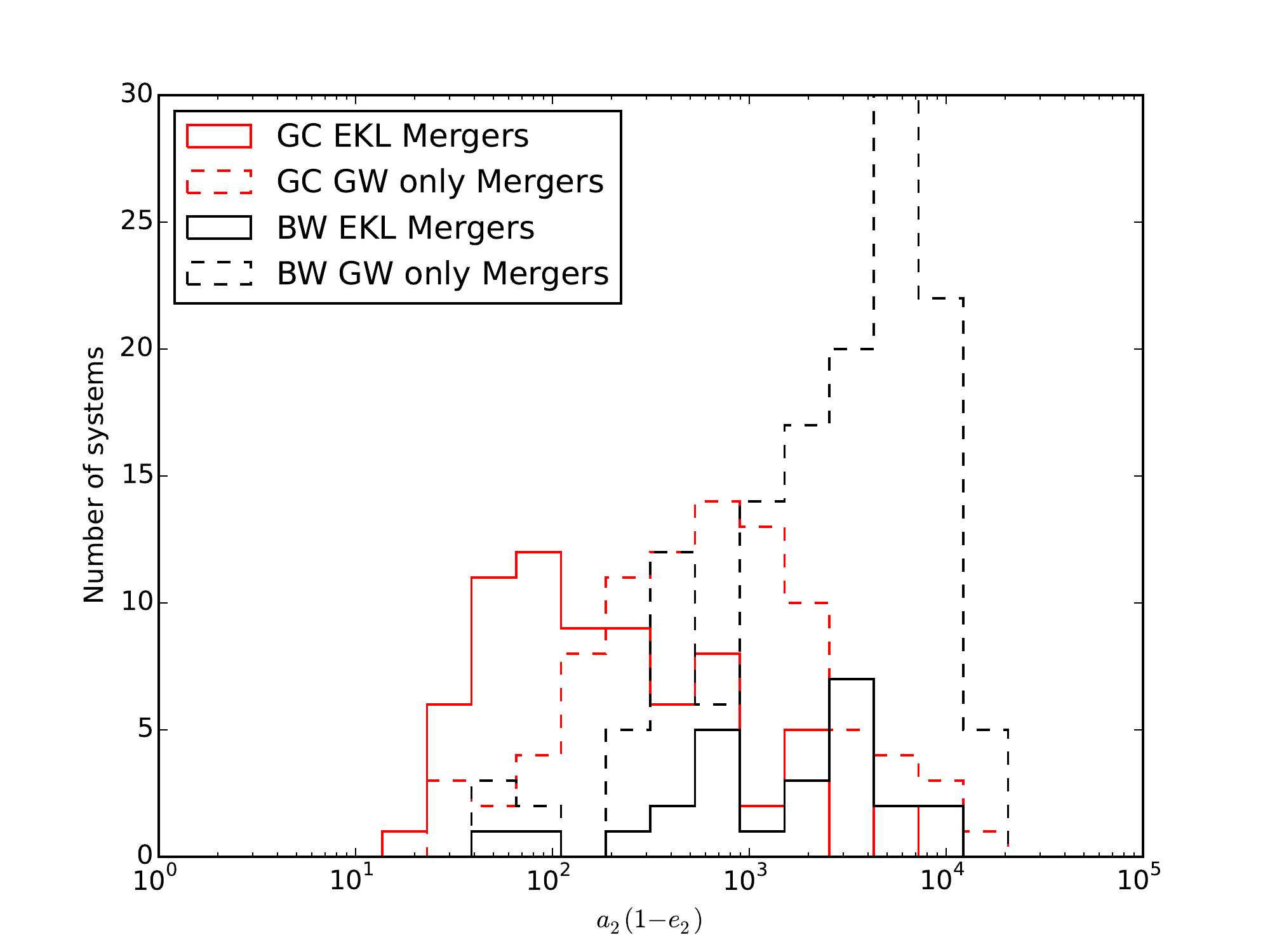}
\end{minipage}
\qquad\caption{{\it Left panel} shows the histogram of $a_2$.  {\it Right panel} shows the histogram of the pericenter $a_2(1-e_2)$. We consider EKL-induced (GW-only) mergers in the BW case in solid (dashed) black lines while the GC mergers are depicted in red lines.   }
\label{fig:a2}
\hspace{1cm}
\end{figure*}

Gravitational perturbations from the MBH can lead to eccentricity excitation of the binary BH which may lead to mergers, via the eccentric Kozai-Lidov (EKL) mechanism \citep[see for review][]{Naoz16}. 
{We integrate the EKL equations \citep[e.g.,][]{Naoz16} including GR precession and GW emission for 1000 (1500) systems in the GC (BW) case either until they merge or until they become unbound, whichever happens first.} The latter takes place on the order of the typical timescale at which close encounters with other stars in the cluster cause the binary to unbind, \citep[e.g.,][]{GalacticDynamics}. This evaporation timescale has the form:
\begin{equation}\label{eq:evap}
t_{\rm ev} = \frac{\sqrt{3}\sigma}{32\sqrt{\pi} G\rho a_1\ln \Lambda}\frac{m_{1}+m_{2}}{m_{3}} \ ,
\end{equation}
where in the inner parts ($\leq 0.1pc$)
\begin{equation}\label{eq:rhoGen}
\rho= \left\{ \begin{array}{rl}
1.35\times 10^6 {\rm M}_\odot ~{\rm pc}^{-3} \left({a_2}/{0.25~{\rm pc}}\right)^{-1.3} & \quad \text{GC} \\  
\frac{3-\alpha}{2\pi}\frac{M_{\rm MBH}}{r^3}\left( {G\sqrt{M_{\rm MBH} M_0}}/{(\sigma_0^2 r)}\right)^{-3+\alpha}  & \quad \alpha=1.5 \hspace{2 mm} \text{BW} \end{array} \right.
\end{equation} 
\citep[see,][for the two cases]{Genzel+2010,Tremaine+02}
is the density of the surrounding stars, $m_1$ and $m_2$ are the masses of the two stellar mass BHs, $m_{3}$ is the average mass of the background stars, $G$ is the universal gravitational constant, $\ln \Lambda$ is the Coulomb logarithm, and $\sigma$ is the velocity dispersion. We adopted $\ln \Lambda = 15$,  $\sigma = 280$~km~s$\rm ^{-1}\sqrt{0.1~pc / a_2}$, and $m_3 = 1~\msun$ \citep{Kocsis+Tremaine}.  For the BW case, $\sigma_0=200$~km~s$^{-1}$ and $M_0 = 3 \times 10^8~\msun$ are constants. Note that the density distributions of the background stars are flatter than the density distribution of the BHs.

{In the GC case, the average evaporation timescale is about $200$~Myr, but there is a very broad distribution from $\sim1$~Myr to $\sim3$~Gyr.  
In the BW case, the average evaporation timescale is about  $120$~Myr, but again there is a broad distribution from $\sim1$~Myr to $\sim2$~Gyr.
Binaries that do not merge are all evaporated by 10 Gyr, consistent with \citet{Stephan+16}.}
 
The above evaporation time assumes that the unbinding is taking place through interaction with the surrounding background stars. However, interaction with background stellar mass black holes may result in different evaporation timescales. If the stellar mass BHs follow the stellar density profile in Equation \ref{eq:rhoGen}, then the evaporation timescale is similar to the evaporation timescale due to interactions with stars, up to a numerical factor that comes from the BH mass. However, if the density profile of the stellar BHs is steeper, like the one that might be expected from mass segregation over long timescales \citep[e.g.][]{2009ApJ...698L..64K}, the evaporation timescale might be shorter. Thus, we have also calculated two representative evaporation timescales resulting from  steeper profile for $10$~$M_\odot$ and $30$~$M_\odot$ black holes, using densities from \citet{Aharon+Perets} for these background black holes (see table 1). These evaporation timescales are much shorter on average than the evaporation timescales resulting from the blackground stars, and thus the absolute number of mergers is reduced. However, the merger rate is not reduced, as explained in Section \ref{sec:additionaltests}.

\section{Results: EKL-induced mergers and GW-only mergers}\label{sec:results}

\subsection{Merging Channels} 
The EKL mechanism has been shown to play an important role in producing short period binaries and merged systems \citep[e.g.,][]{Tho10,Naoz+11,Antonini+12,Prodan+13,NF,Prodan+2015,Antonini+16,Stephan+16,Naoz+16,Naoz16}. The high eccentricity values achieved during the binary evolution leads to a shorter GW emission timescale, which may cause a merger before the binary become unbound as shown in Figure \ref{fig:OctvsQuad}. If a merger takes place due to high eccentricity excitation we denote this merger as an {\bf EKL-induced  merger}. 

 The eccentricity excitation due to EKL can be inhibited by GR precession if the timescale for the latter is much shorter than the former \citep[e.g.,][]{Naoz+13GR}. The EKL timescale at the quadrupole level of approximation is estimated as:
\begin{equation}\label{eq:tquad}
t_{\rm quad}\sim\frac{16}{30\pi} \frac{m_1+m_2+m_3}{m_3}\frac{P_2^2}{P_1}(1-e_2^2)^{3/2}  \ ,
\end{equation}
 \citep[e.g.,][]{Antognini15}, where $P$ denotes the orbital period and the GR precession of the inner orbit timescale is:
 \begin{equation}\label{eq:tGR}
t_{\rm GR,inner} \sim 2\pi \frac{a_1^{5/2} c^2 (1-e_1^2) }{ 3 G^{3/2} (m_1+m_2)^{3/2}} \ ,
\end{equation}
\citep[e.g.,][]{Naoz+13GR} where $c$ is the speed of light. 
If $t_{\rm GR,inner} < t_{\rm quad}$, then EKL effects are negligible, the binary evolves due to close encounters with other stars in the cluster, and the inspiral is caused only by GW emission. The binary may approach merger if the GW timescale, $t_{\rm GW}$, \citep{Peters64} is shorter than the timescale it takes the binary to become unbound (see Equation (\ref{eq:evap})). We identify those as {\bf GW-only mergers}. 
 
 In other words we identify two channels for mergers. In the GW-only merger we have $t_{\rm GR}<t_{\rm quad}$ and $t_{\rm ev}>t_{\rm GW}$, and in the EKL-induced merger $t_{\rm quad} < t_{\rm GR}$, in which the eccentricity can be excited to near unity.

\begin{figure*}
\hspace{-3cm}
\centering
\begin{minipage}[b]{.34\textwidth}
\centering
\includegraphics[scale=.45]{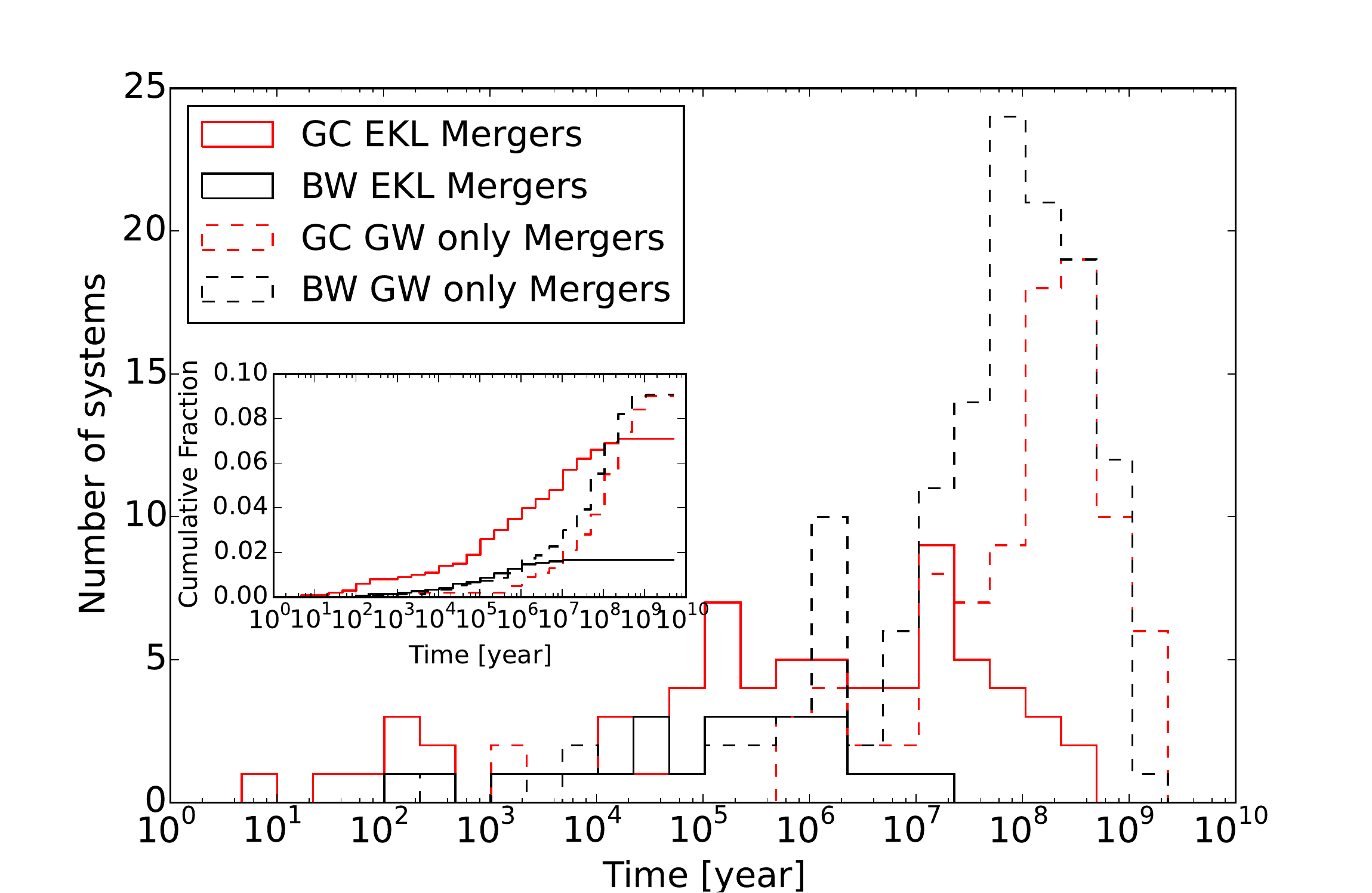}
\end{minipage}\hspace{4cm}
\begin{minipage}[b]{.34\textwidth}
\centering
\includegraphics[scale=.45]{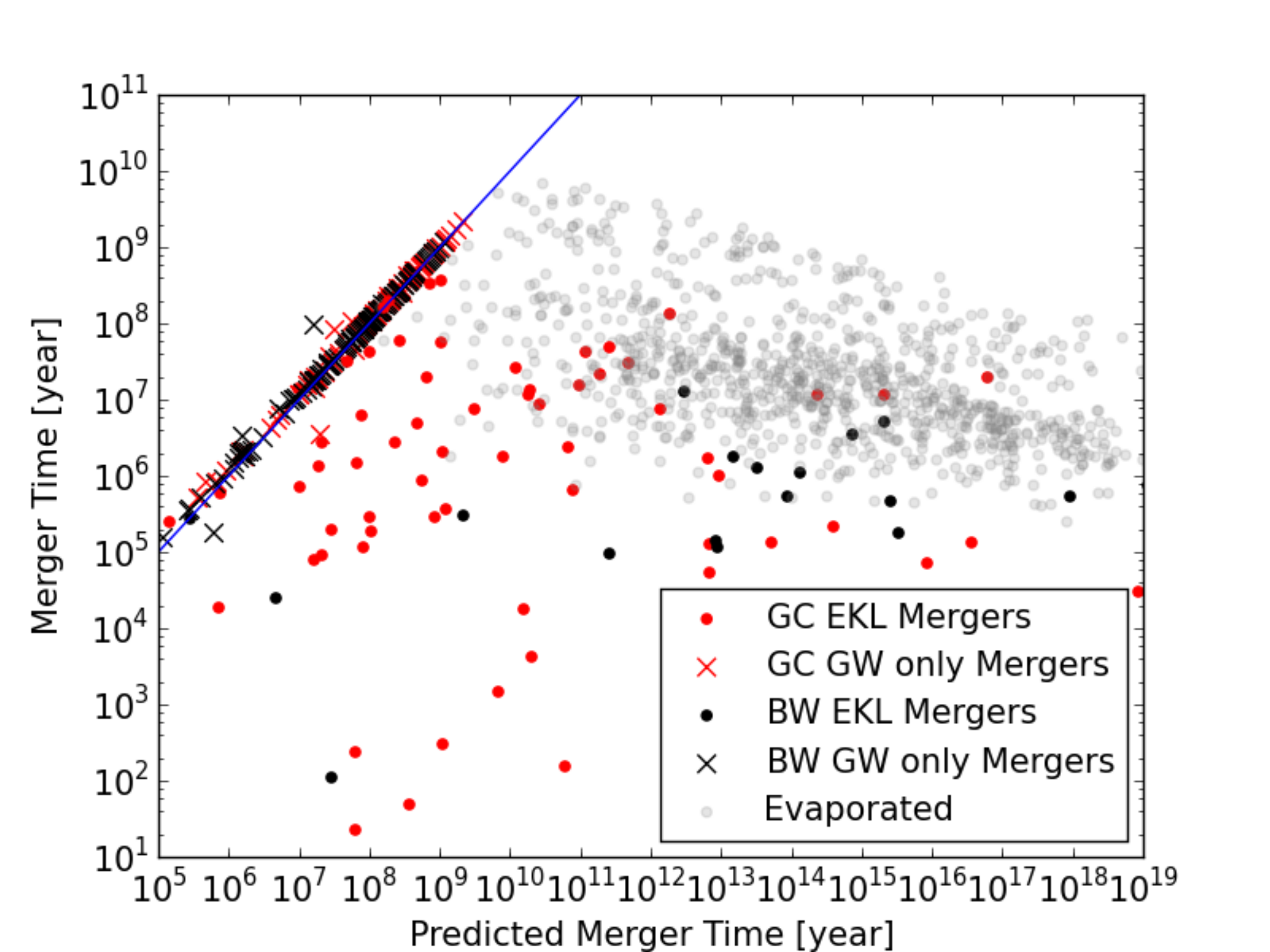}
\end{minipage}
\qquad\caption{ \upshape {\bf GW timescale comparison between different channels.} {\it Left panel:} The histogram of merger times for EKL mergers (red lines) and GW-only mergers (blue lines). The {\bf inset} shows the cumulative distribution function of merger times. {\it Right panel:} The actual merger timescale as a function of the GW timescale given the initial condition of the simulation. The GW-only mergers (blue points) follow the expected $1:1$ line, while most of the EKL-induced merger times are shorter. The gray points never merge and their y-axis value marks their unbinding timescale.}
\label{fig:tGW}
\hspace{1cm}
\end{figure*}

The EKL-induced high eccentricity excitations usually appear in a distinctive regime in the $\epsilon-i$ parameter space. This can be seen in Figure \ref{fig:epsi}, where the EKL-induced mergers inhabit a specific regime of higher $\epsilon$ (between $10^{-4}$ and $10^{-2}$) and relatively large inclinations \footnote{Although we note, that some EKL-induced mergers  take place beyond the nominal Kozai angles, (i.e., $i<40^\circ$ and $i>140^\circ$), which is consistent with the near co-planar behavior \citep{Gongjie+14}.}. The EKL-induced mergers represent $\sim 7\%$ ($1.7\%$) of all GC (BW) Monte-Carlo systems. The systems that merge via the GW-only process (where no significant eccentricity excitations took place) occupy this parameter space uniformly. 
They represent about $9\%$ ($9.1\%$)  of all the GC (BW) systems. The two sub-panels in Figure \ref{fig:epsi} show that the EKL yields a systematically shorter merger timescale. {In other words, out of all systems that merged,  $44\%$ are EKL-induced mergers in the GC case and  about $16\%$ are EKL-induced mergers for the BW density profile.}

As implied from Figure \ref{fig:epsi} these two merger channels can yield different predictions for the statistics of BH binary mergers. The EKL-induced mergers take place preferentially in systems that are closer to the MBH (see left panel of Figure \ref{fig:a2}), whereas the GW-only systems have no apparent trend. We note that the cut-off in GW-only mergers for  $a_2<200$~AU takes place because systems below this value either have GR precession timescale much longer than the EKL, or unbind before the binary can merge.

The EKL-induced mergers systematically happen on shorter timescales compared to the GW-only mergers (see inset on Figure \ref{fig:tGW}).
On average EKL-induced mergers in the GC case take place within about $25$~Myrs while GW-only mergers merge on average after $296$~Myrs (see Figure \ref{fig:tGW} for the merger time histogram of the two channels).
On the other hand, the BW case yields a shorter unbinding timescale and thus we get that the average EKL-induced mergers in the BW is about 1 Myr while the GW-only merger is about $166$~Myr.  

{As expected, in both of our examples, none of the EKL-induced mergers were initially hard.  On the other hand, out of the GW-only mergers, $14\%$ in the GC example and $43\%$ of the BW example, were, in fact, hard binaries initially. Hardening channels \citep[e.g.,][]{Quinlan96} may result in a shorter merger time for those binaries.   }

\begin{figure*}
\hspace{-3cm}
\centering
\begin{minipage}[b]{.4\textwidth}
\centering
\includegraphics[scale=.45]{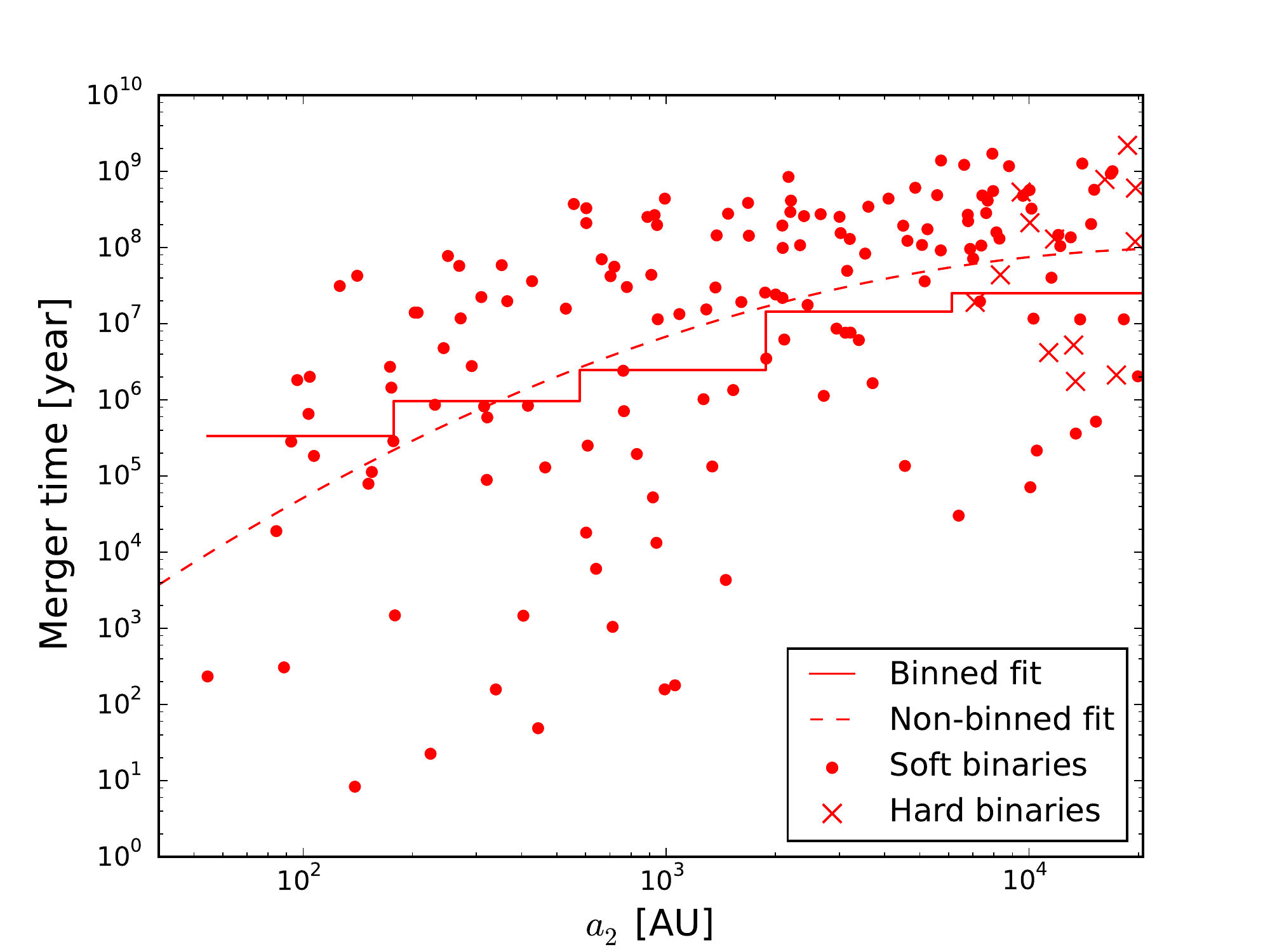}
\end{minipage}\hspace{1cm}
\begin{minipage}[b]{.4\textwidth}
\centering
\includegraphics[scale=.45]{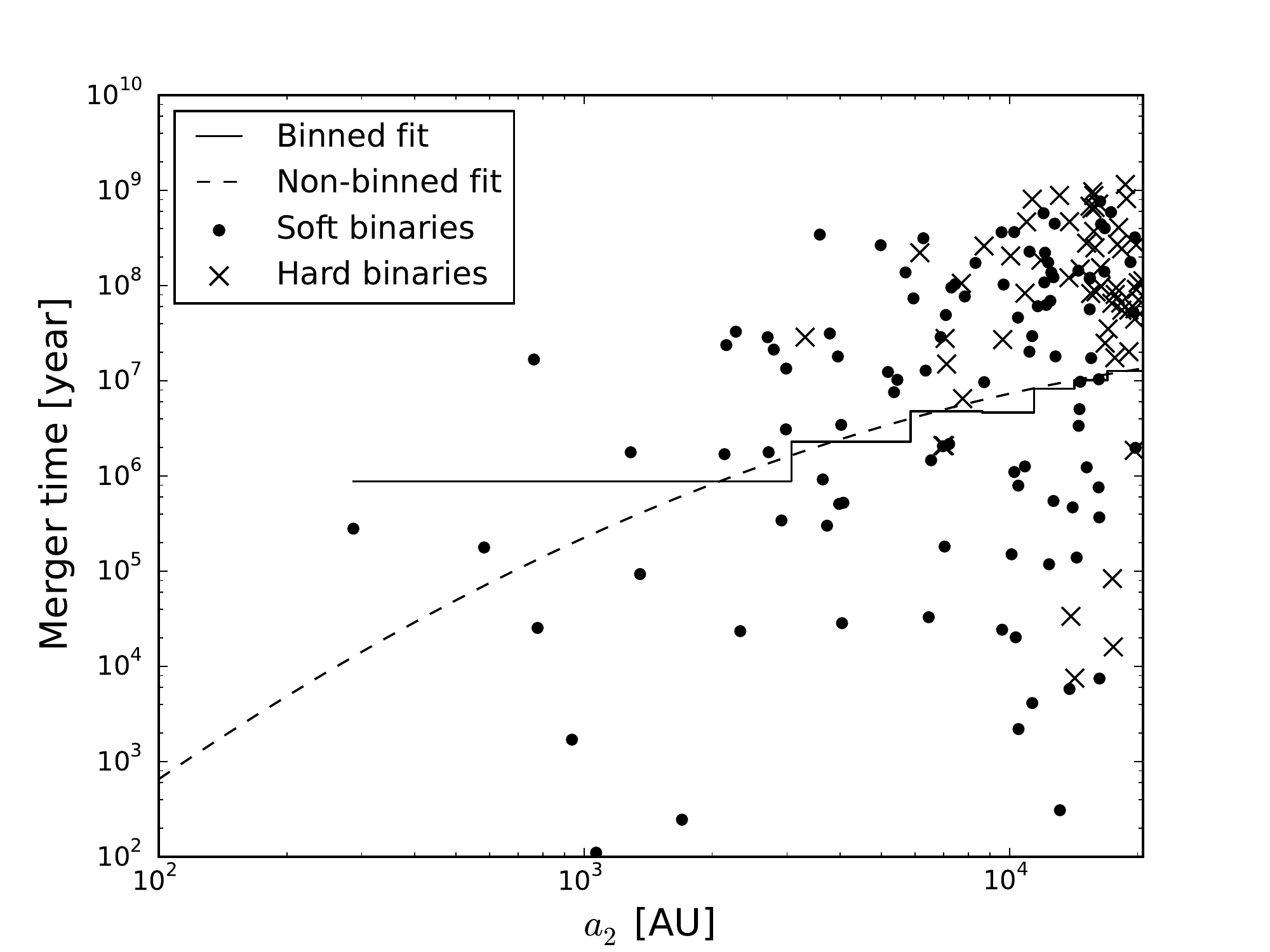}
\end{minipage}
\qquad\caption{Merger time of soft and hard binaries as a function of outer binary separation $a_{2}$ for the GC case ({\it left panel}), and the BW case ({\it right panel}). We perform both a binned (solid line) and not binned (dashed line) bootstrap fit to the data and find a slight dependence of merger time on $a_{2}$.
\label{fig:tmerge_v_a2}}
\hspace{1cm}
\end{figure*}

\begin{figure*}
\hspace{-3cm}
\centering
\begin{minipage}[b]{.4\textwidth}
\centering
\includegraphics[scale=.45]{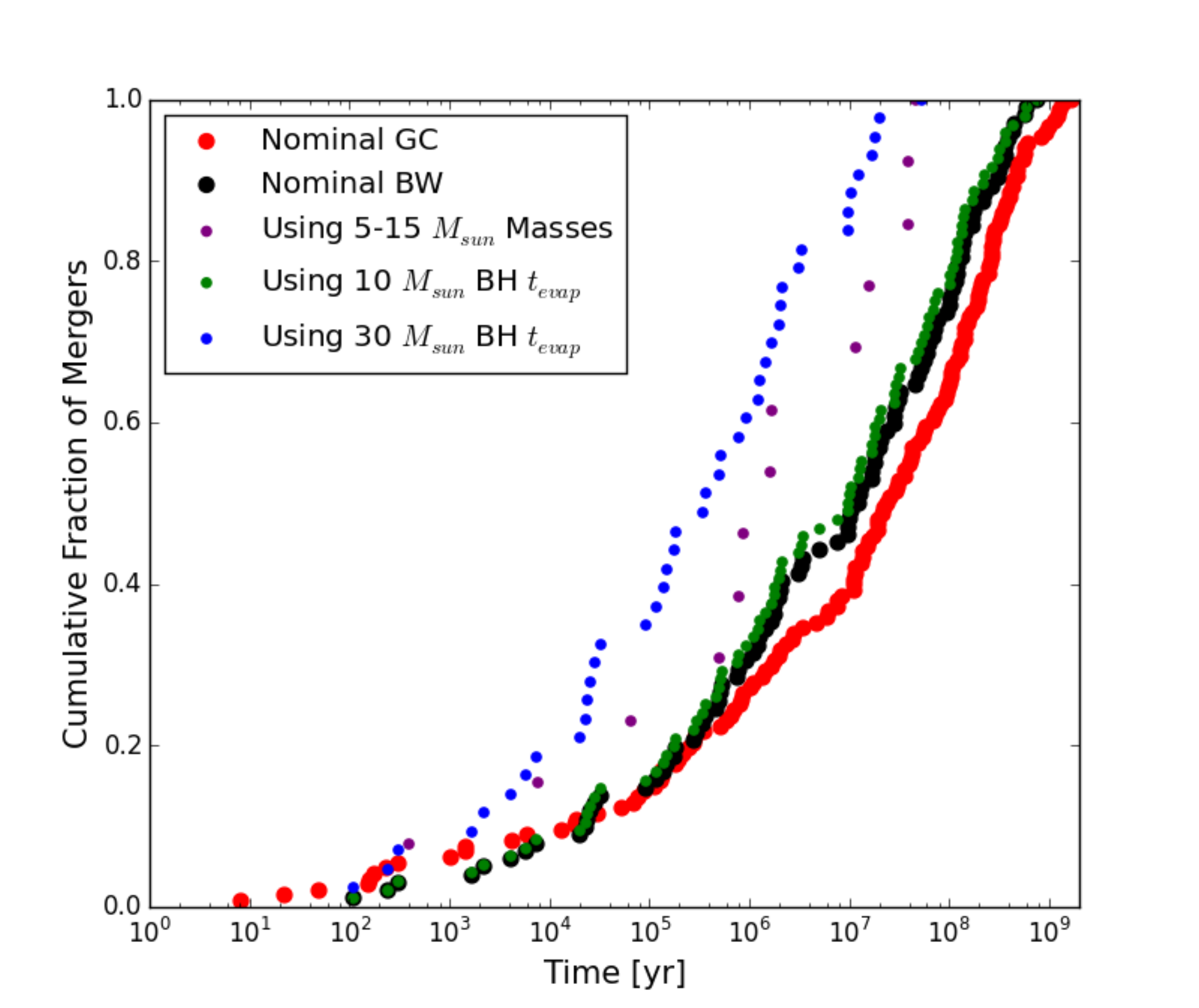}
\end{minipage}\hspace{1.2cm}
\begin{minipage}[b]{.4\textwidth}
\centering
\includegraphics[scale=.45]{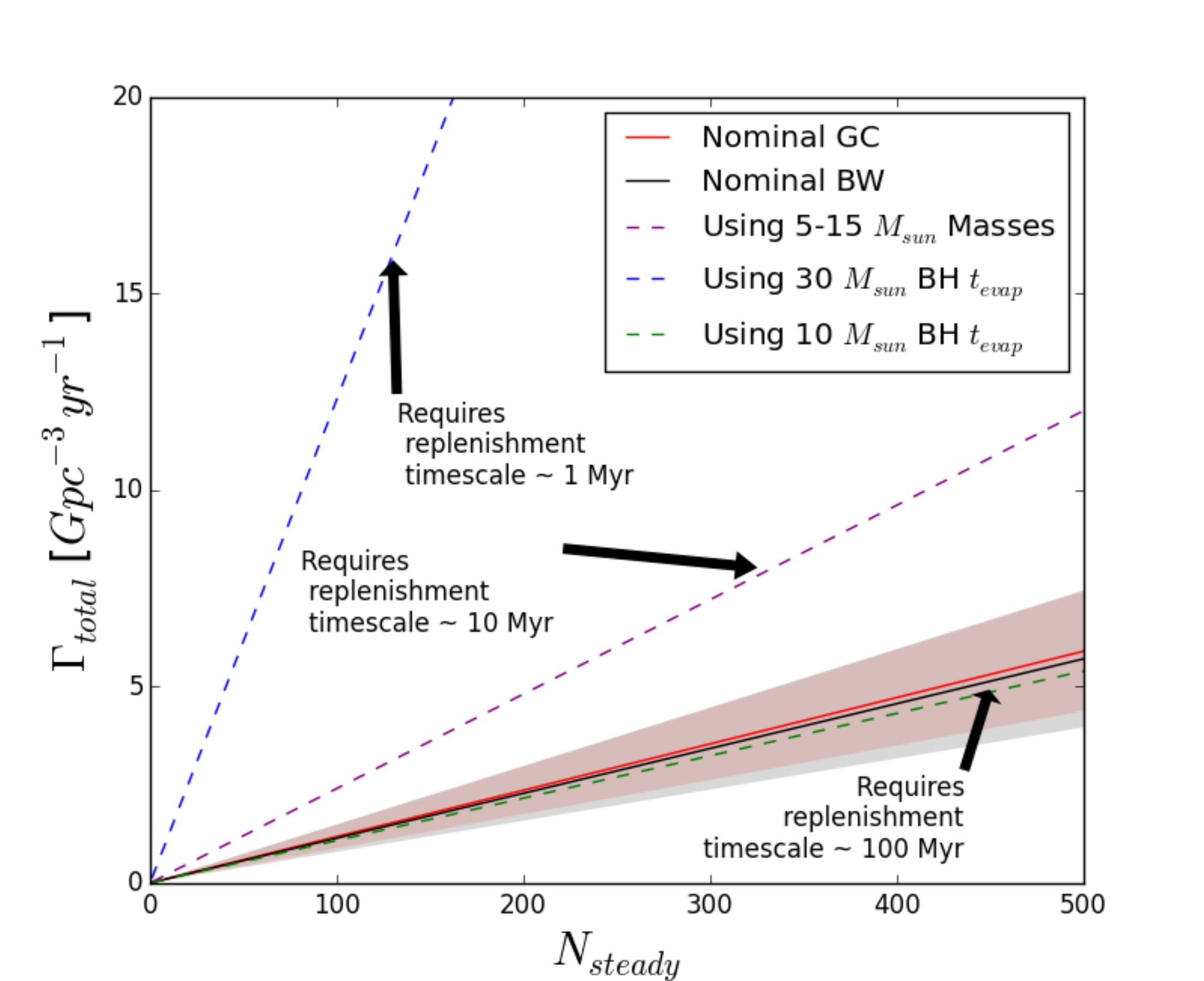}
\end{minipage}
\qquad\caption{{\it Left panel:} Scatter points show the cumulative fraction of mergers as a function of time for the GC case (red), the BW case (black), using a 5-15 $\msun$ mass distribution (purple), using an evaporation timescale dominated by 10 $\msun$ black holes (green), and using an evaporation timescale dominated by 30 $\msun$ black holes (blue). Cases that take a shorter time to reach a cumulative merger fraction of unity have a shorter average merger time and therefore a higher total merger rate, shown in the right panel. {\it Right panel:} The total merger rate $\Gamma_{\rm total}$ as a function of the number of steady state BH-BH binaries in the galactic nucleus $N_{\rm steady}$ for the various cases in the left panel. The shaded areas show upper and lower limits to the GC and BW rates corresponding to the bootstrap distribution. Also shown are the approximate BH binary replenishment timescales required to maintain the merger rates shown. The two shorter replenishment timescales 10 Myr, and 1 Myr, are probably unlikely to take place. Thus, the rate represented by the nominal GC and BW cases, requiring a replenishment timescale of 100 Myr, is the most likely scenario.}
\label{fig:GammaDis}
\hspace{1cm}
\end{figure*}

\subsection{Rate Estimate}
We also estimate the merger rate of BH-BH systems. The total merger rate is dominated by the merger rate of the soft binaries because the hard binaries represent a very small percentage of the total number of systems in our simulation ($2\%$ for the GC case and $5\%$ for the BW case). Thus, we only perform the following calculations for the soft binaries.The merger rate per unit volume is defined as:
\begin{equation}\label{eq:rate}
\Gamma_{\rm total} = n_{g} \Gamma f_{MBH}  \ ,
\end{equation}
where $n_g$ is the density of galaxies, $\Gamma$ is the merger rate per galaxy, and $f_{\rm MBH}$ is the fraction of galaxies containing a MBH.  We adopt  $n_g = 0.02\Mpc^{-3}$ \citep{2005ApJ...620..564C} and  $f_{\rm MBH} = 0.5$ following \citep{2015ApJ...806L...8A,2015ApJ...812...72A}. Note that this is a rather conservative number and we expect this fraction to be higher. 

The merger time for our systems span a very broad range of timescales with a slight systematic trend for shorter merger time closer to the MBH (i.e., smaller $a_2$) as can be seen in Figure \ref{fig:tmerge_v_a2}. However, these short-lived systems represent a small fraction of the total number of merged systems, and the apparent trend may be a result of statistical bias. To ascertain whether the merger rate has a dependency on $a_2$, we perform a bootstrap resampling of the merger time versus the $a_2$ distribution. We find that a bootstrapped analysis (both binned and not binned in $a_2$) yields a dependency in $a_2$ (see Figure \ref{fig:tmerge_v_a2}). However, the percentage of systems close to the SMBH is small. Thus, we proceed by calculating the cumulative fraction of mergers as a function of time to calculate an average merger rate.

Figure \ref{fig:GammaDis} shows the cumulative fraction of merged systems as a function of time for various cases. We find that the cumulative {\it number} of merged systems fits a power law of the form:
\begin{equation}\label{eq:fitlaw}
N_{merged} = a~t^b \ ,
\end{equation}
where $a$ and $b$ are constants. We perform a bootstrap resampling on the cumulative number of mergers versus merger time distribution and generated 500 resampled data sets for the GC and BW cases. We fit the above power law to all of these data sets, which gives us a range of possible fits. The average best fit values for $a$ and $b$ are 2.8 and 0.19 for the GC case, and 2.3 and 0.19 for the BW case. We note that we also fit a combination of different exponential functions to the cumulative number of mergers (not shown), which perhaps follows a exponential decay-like process. These fits were systematically less good, but nonetheless yield consistent results with those shown below. We can define a "half-life", $t_{1/2}$, for each sample as the time it takes for half of the systems that will merge to merge. Thus, the merger frequency at $t = t_{1/2}$ is:
\begin{equation}\label{eq:mergefrequency}
\gamma = \frac{1}{N}\frac{dN}{dt} \Big|_{t=t_{1/2}} = \frac{b}{t_{1/2}} \ .
\end{equation}
We use this frequency as the characteristic merger frequency for each data set. 

\begin{table}\label{table:avetimes}
\begin{center}
\begin{tabular}{| c | c | c | c |}
\hline
{\bf Case} & \bm{$\langle t_{\rm evap} \rangle$ } {\bf [yr]} &  \bm{$1/\gamma$} {\bf [yr]} & \bm{$f_{\rm merge}$} \\ [0.5ex]
\hline \hline
{\bf Nominal GC} & $2.1~\times~10^8$  & $1.3~\times~10^8$ & 0.15 \\
\hline
{\bf Nominal BW} &  $1.2~\times~10^8$ & $6.3~\times~10^7$ & 0.07 \\
\hline
{\bf BW, 5-15} {$\msun$} {\bf BBH} & $7.7~\times~10^7$ & $1.1~\times~10^7$ & 0.03\\
\hline
{\bf BW, 10}  {$\msun$} {\bf BH} \bm{$t_{\rm evap}$} & $1.0~\times~10^8$  &   $6.3~\times~10^7$ & 0.07\\
\hline
{\bf BW, 30} {$\msun$} {\bf BH} \bm{$t_{\rm evap}$}  & $5.9~\times~10^6$ & $2.5~ \times~10^6$  & 0.03  \\
\hline
\end{tabular}
\caption{The table above lists the average evaporation timescale, the inverse of the merger frequency $\gamma$ as defined in Equation \ref{eq:mergefrequency}, and the merger fraction for each of the cases shown in Figure \ref{fig:GammaDis}. As can be seen from the table, a shorter average evaporation timescale means a shorter average merger timescale ($1/\gamma$), and a smaller merger fraction $f_{\rm merge}$. The binary replenishment timescale required for steady state is roughly on the same order of magnitude as the average evaporation and merger timescales.}
\end{center}
\end{table}

Assuming a steady state number of BH binaries in the galactic nucleus, $N_{\rm steady}$, we can then estimate the merger rate per galaxy $\Gamma$ as:
\begin{equation}
\Gamma = N_{\rm steady} f_{\rm merge} \gamma \ ,
\end{equation}
where $f_{\rm merge}$ is the merger fraction from our simulation, equal to 0.15 for the GC case and 0.07 for the BW case. $N_{\rm steady}$ is highly uncertain, so we set it to be a free parameter between 1 and 500.  The right panel of Figure \ref{fig:GammaDis} shows the total merger rate as a function of $N_{\rm steady}$. 
We use a nominal value of $N_{\rm steady} ~=~ 200$ to calculate the merger rates for the GC and BW cases. We find a nominal average merger rate of $\sim$ 2 $\Gpc^{-3} \yr^{-1}$ for both the GC and BW cases. The average merger rate values are represented by the red and black solid lines in the right panel of Figure \ref{fig:GammaDis}; the merger rate range obtained from the bootstrap is depicted by the shaded areas.

Our estimated merger rates for the GC and BW cases are on the same order as the merger rates estimated for globular clusters, $5$~Gpc$^{-3}$~yr$^{-1}$ \citep{Rodriguez+2016}, isolated triples in galaxies 0.14--$6\Gpc^{-3}\yr^{-1}$ \citep{2016arXiv160807642S}, and mergers following close encounters of initially unbound BHs 0.04--$3\Gpc^{-3}\yr^{-1}$ \citep{OKL09}\footnote{Here we use the rates per single galaxy in \citet{OKL09} Table 1 and multiply by $\xi=10$ and $n_g=0.02\Mpc^{-3}$. However, that mechanism is weakly sensitive to the NSC and MBH mass. Dwarf galaxies dominate the rates with a much higher $n_g$.}. 
The current LIGO/VIRGO detection constrain the total merger rate of circular BH binaries to within 12--$240\Gpc^{-3}\yr^{-1}$ \citep{LIGO3}. 

Note that \citet{Antonini+12} rate estimation is based on the quadrupole based semi-analytical timescale which does not capture the system's full dynamical behavior. The octupole level of approximation, used here, shortens the merger timescale for $16\%-40\%$ of the EKL-induced mergers, for the GC and BW distributions, respectively (see Figure \ref{fig:OctvsQuad}). Furthermore, \citet{Antonini+12}  used the semi-analytical timescale given in \citet{Tho10} to estimate the merger timescale. However, this does not capture the correct merger timescale for each binary BH as can be deduced from the inset in Figure  \ref{fig:OctvsQuad}. 

\subsection{Additional Tests}\label{sec:additionaltests}
As stated in Section \ref{sec:ICs}, we have also considered the effects of using a narrower black hole mass distribution ranging from $5-15$~$M_\odot$ instead of $6-100$~$M_\odot$ for the BW case, while keeping all other parameters the same; and the effects of using the evaporation timescale resulting from interactions with $10$~$M_\odot$ and $30$~$M_\odot$ background black holes instead of background stars. For the $5-15$~$M_\odot$ black hole mass distribution, the average evaporation time is shorter than the nominal GC and BW cases because of the reduced BH binary mass (see Equation \ref{eq:evap} and Table \ref{table:avetimes}). We performed 500 Monte Carlo simulations and found that the number of GW-only mergers are greatly reduced, due to the fact that they generally take longer to merge. However, the short-timescale EKL mergers are unaffected, and we find that using a $5-15$~$M_\odot$ black hole mass distribution results in a merger population that predominantly consists of short-timescale EKL mergers. We also run a set quadrupole-only simulation for the EKL mergers in this case, and found that about $\sim25\%$ take longer to merge than with the octupole, or do not merge at all (the total percentage of non-mergers with only quadrupole are $\sim 13\%$). Fitting Equation (\ref{eq:fitlaw}) to the number of mergers as a function of time results in a shorter merger half-life (this can be seen from the left panel of Figure \ref{fig:GammaDis}), and therefore a higher merger frequency as defined in Equation (\ref{eq:mergefrequency}). Thus, performing the calculation for $\Gamma_{\rm tot}$ results in a merger rate that is {\it higher} than before, as can be seen in the right panel of Figure \ref{fig:GammaDis}. This makes sense, considering that a majority of the mergers in this case are EKL mergers that take place very quickly. However, a shorter average evaporation timescale and average merger timescale for a fixed steady-state number of binaries also means that we require a quicker BH binary replenishment mechanism (on a timescale roughly equal to the average evaporation and average merger timescale) to maintain steady state. Otherwise the high rate of merger seen in Figure \ref{fig:GammaDis} cannot be maintained. For this particular case, a replenishment timescale of about $\rm 10~Myr$ is required to maintain steady state.

We have a similar consequence resulting from using the evaporation timescales from interactions with background black holes of masses $10$~$M_\odot$ and $30$~$M_\odot$ rather than background stars. To do this, we first calculate the new evaporation timescales using the number densities for $10$ $M_\odot$ and $30$ $M_\odot$ found in \citet{Aharon+Perets}, then we compare the merger time of each of our soft binaries with the new evaporation timescales, and if the merger time is longer, we count it as a non-merger. We then fit the new merger distributions as a function of time to Equation (\ref{eq:fitlaw}) as done previously, and then calculate $\Gamma_{\rm tot}$ as a function of $N_{\rm steady}$, shown in the right panel of Figure \ref{fig:GammaDis}. We find that using the evaporation timescale dominated by the $10$~$M_\odot$ black holes results in a merger rate very similar to the nominal BW and GC cases, but using the evaporation timescale dominated by the $30$~$M_\odot$ black holes results in a much higher merger rate than before (see right panel of Figure \ref{fig:GammaDis}). In the $30$~$M_\odot$ background black holes case, the average evaporation timescale is very short compared to the nominal GC and BW cases (see Table \ref{table:avetimes}). Similar to what happened with the  $5-15$~$M_\odot$ black hole mass distribution, only BH binaries with short merger timescales can merge before they are evaporated, so the average merger timescale is very short, leading to a higher $\Gamma_{\rm tot}$. Furthermore, like in the $5-15$~$M_\odot$ black hole mass distribution case, this higher $\Gamma_{tot}$ will require a shorter binary replenishment timescale (roughly one Myr) to continue in steady state with a fixed number of BHs. For the $10$~$M_\odot$ background black holes case, the average evaporation timescale is very similar to the the average evaporation timescale in the nominal GC and BW cases (see Table \ref{table:avetimes}). Thus, the average merger timescale is very similar to before, leading to a similar $\Gamma_{\rm tot}$. Note that since the evaporation timescale has a weak dependence on perturber mass, the significantly shorter evaporation timescale resulting from the $30$~$M_\odot$ is primarily consequence of their steeper density profile as compared to the $10$~$M_\odot$ background black holes.

Finally, we have also re-run our EKL mergers run in the GC case with Newtonian precession added. First, one should differentiate between Newtonian precession that occurs due to a spherical mass distribution (which was what we assumed in the paper), versus precession due to deviations from spherical symmetry as explored in \citet{Petrovich17}. Thus, in the following analysis we constrain ourselves to the spherically symmetric Newtonian precession. The Newtonian precession in our case causes the outer orbit to precess. While it may take place on similar timescale as the octupole timescale, it will not affect the inner orbit precession, which has a much greater effect on the EKL mechanism. However, to quantify how much Newtonian precession will affect our systems, we re-run the 63 EKL mergers for the GC case with Newtonian precession included, following Equation (44) in Tremaine (2005). We found that over 90\% of these runs remained mergers even with Newtonian precession included. Furthermore, the merger time distribution remains the same.

\section{Discussion}\label{sec:dis}

We investigated the secular evolution of stellar-mass BH binaries in the neighborhood of the MBH in galactic nuclei using  Monte Carlo simulations. Our equations included the hierarchical secular effects up to the octupole-level of approximation \citep[the so-called EKL mechanism,][]{Naoz16}, and general relativity precession of the inner and outer orbits, and gravitational wave emission between the two stellar BHs. During their evolution, binary stellar mass BHs may undergo large eccentricity excitations which can drive them to merge (see for example Figure \ref{fig:OctvsQuad}). 
 While BHs are expected to segregate toward the center of the galactic nuclei, the power law index of the number density of the cusp is still uncertain. As a proof-of-concept, we explored two cases: $r^{-2}$, denoted as BW; and $r^{-3}$, denoted as GC. We find a consistent merger rate in both cases.

We identified two channels for BH-BH mergers. In the first type of merger, the general relativity precession timescale is shorter than the quadrupole timescale, suppressing eccentricity excitations. However, the initial GW merger timescale is shorter than the evaporation timescale, leading to a merger. We call these GW-only mergers. These systems will merge in the absence of an MBH in the center of a galaxy. For the second type of merger, the EKL mechanism produces large eccentricity oscillations in the inner binary orbit, driving the binary BHs to merge. We call these EKL-induced mergers.

GW-only and EKL-induced mergers occupy different parts of the parameter space for both number density profiles we examined. Specifically, considering the $\epsilon-i$ parameter space, where $\epsilon$ is given by Eq.~(\ref{eq:epsilon}).  EKL-induced mergers will preferentially occupy a regime of high $\epsilon$ and close to $90^\circ$ mutual inclination. On the other hand, GW-only  mergers are uniformly spread in $i$ and $\rm ln(\epsilon)$ (see Figure \ref{fig:epsi}). This yields a prediction for the merger distribution as a function of distance from the MBH: the EKL-induced mergers will be systematically closer to the MBH. This is depicted in Figure  \ref{fig:a2}.

Finally we estimate the total merger rate to be at the order of unity $\Gpc^{-3} \yr^{-1}$ and perhaps even higher. This rate is coincidentally comparable to the estimated merger rate in globular clusters, which suggests that galactic nuclei may host a significant fraction of the BH-BH mergers. Note that \citet{Antonini+16} showed that if the natal kick of BH binaries is higher than $50$~km~sec$^{-1}$, then the efficiency of forming BH binaries in globular clusters is suppressed significantly \citep{Chatterjee+2016}\footnote{And also above this natal kick the mergers of isolated binaries is significantly suppressed \citep{Belczynski+2016} }, and BH mergers in nuclear star clusters dominate over mergers in globular clusters. 
We note that, if natal kicks are smaller, so that BH binaries form efficiently in globular clusters, BH mergers in nuclear star cluster with an MBH is still comparable to that of globular clusters.

We note that  we have neglected the effects of resonant relaxation, as they should be added self-consistently \citep{Rauch+Tremaine,Sridhar+Touma}. These effects operate on longer timescales than the quadrupole timescale. However, they can refill the high inclination EKL regime in some cases, causing further eccentricity excitation. This is beyond the scope of this paper. Recently \citet{Petrovich17} showed that variations from spherical symmetry in the potential may induce eccentricity excitation at large $a_2$. Thus, their proposed mechanism may even increase the merger rate discussed here for systems at larger distances. However, their study also neglected vector resonant relaxation which operates on the same timescales and distances. Therefore, further investigation is needed for a more accurate treatment of these effects. 

Our results suggest that the key parameter controlling the merger rate is the steady-state number of BH-BH binaries. We find that, for typical BH--BH binary populations often assumed in the literature, the predicted merger rate is $\sim 1 - 3$$\Gpc^{-3} \yr^{-1}$; but this can become much higher, depending on the exact assumptions made (see Figure \ref{fig:GammaDis}).
Our steady-state assumption requires a binary replenishment mechanism, without which the EKL-induced BH-BH merger rate would rapidly decay (as implied from Figure \ref{fig:tGW}). The BH-BH binary population in real systems may be replenished from the outside, for example from globular clusters that spiral in, or from nearby star formation \citep[e.g.,][]{2006ApJ...645.1152H,2006ApJ...645L.133H,Gnedin14,2015ApJ...812...72A,Aharon+Perets}. A different channel for recent in-situ formation of BH--BH binaries may exist near the centers of E+A, or post-starburst, galaxies \citep[e.g.,][]{Dressler+83}. These galaxies are special because it appears that they underwent a star formation episode that terminated abruptly $\sim 1$~Gyr ago. Thus, this population may hold recent BH-BH binaries near their nuclei. While these galaxies are a relatively rare subtype of elliptical galaxies, they may increase the stellar mass of the galaxy by $\sim10\%$ \citep[e.g.,][]{Swinbank+12}. Thus, due to their recent star formation and overdensity in the nuclei, these galaxies were linked to an enhancement of tidal disruption events \citep[e.g.,][]{Arcavi+14,Stone+16}. Therefore, these galaxies may have an overabundance of BH binaries, which can cause enhancement of BH-BH mergers. Recently, \citet{Bartos+2017} showed that with a big enough sample of observed BH-BH mergers, a certain BH-BH merger channel can be statistically correlated with rare galaxy types, including E+A galaxies. This may prove a valuable test for our model in the future.


The BH-BH merger scenario presented here, in our proof-of-concept examples, suggests that MBH gravitational perturbations can enhance the merger rate. Since these mergers take place close to the SMBH, the acceleration of the binary around the SMBH may cause a detectable Doppler shift of the waveform \citep{Meiron+17,Inayoshi+17}. Furthermore, a GW echo of the binary BH merger caused by the SMBH may be detectable \citep{Kocsis2013}. We have provided the parameter space at which these mergers take place, and demonstrated that they occupy a different part of the parameter space. This suggests that this channel may be statistically distinguished with a sufficiently large sample of mergers (Hoang et al. in prep.).

Finally, motivated by the calculation of the rate of BH-BH mergers for $>0.1$~pc by \citet{Petrovich17}, we follow their set of assumptions of a compact object formation rate between $2 \times 10^{-5} - 10^{-4}$~$\rm yr^{-1}$, with a fraction of surviving BH-BH binaries of $2.5 - 4.5\%$. The merger fraction from \citet{Petrovich17} along with these assumptions yields a merger rate of $0.6-15$ $\Gpc^{-3} \yr^{-1}$. Similarly, with our merger fraction and these assumptions we find a merger rate of $0.7-14$ $\Gpc^{-3} \yr^{-1}$ for our $< 0.1$~pc regime. Therefore, assuming that these assumptions are correct, we conclude that the BH-BH merger rate in the proximity of galactic nuclei can be as significant as $\sim 30$ $\Gpc^{-3} \yr^{-1}$.

\acknowledgments{
We thank Michael Fitzgerald for insightful discussions about statistical analysis.
BMH  acknowledges the support of the Eugene V. Cota-Robles Fellowship and the Graduate Dean's Scholar Award. 
SN acknowledges partial support from a Sloan Foundation Fellowship.
FAR and FD acknowledge support from NASA Grant NNX14AP92G and NSF Grant AST-1716762 at Northwestern University.
This project has received funding from the European Research Council (ERC) under the European Union's Horizon 2020 research and innovation programme under grant agreement No 638435 (GalNUC) and by the Hungarian National Research, Development, and Innovation Office grant NKFIH KH-125675.}

\bibliographystyle{yahapj}

\end{document}